\documentclass[apjl,a4paper,12pt,useAMS]{emulateapj}
\usepackage{amsmath}
\usepackage{cases}
\usepackage{amssymb}
\usepackage{graphicx}
\usepackage{natbib}

\begin{document}

\title{Optical and radio transients after the collapse of super-Chandrasekhar white dwarf merger remnants}
\author{Yun-Wei~Yu$^{1,2}$, Aming Chen$^{1,2}$, and Bo Wang$^{3}$}

\altaffiltext{1}{Institute of Astrophysics, Central China Normal
University, Wuhan 430079, China, {yuyw@mail.ccnu.edu.cn}}
\altaffiltext{2}{Key Laboratory of Quark and Lepton Physics (Central
China Normal University), Ministry of Education, Wuhan 430079,
China}
\altaffiltext{3}{Key Laboratory for the Structure and Evolution of Celestial Objects, Yunnan Observatories, CAS, Kunming 650216, China }

\begin{abstract}
Super-Chandrasekhar remnants of double white dwarf mergers could sometimes collapse into a rapidly rotating neutron star (NS), accompanying with a mass ejection of a few times $0.01M_{\odot}$. Bright optical transient emission can be produced by the ejecta due to heating by radioactivities and particularly by energy injection from the NS. Since the merger remnants before collapse resemble a star evolving from the asymptotic giant branch phase to the planetary nebula phase, an intense dusty wind is considered to be driven about several thousand years ago before the collapse and surround the remnant at large radii. Therefore, the optical transient emission can be somewhat absorbed and scattered by the dusty wind, which can suppress the peak emission and cause a scattering plateau in optical light curves. Several years later, as the ejecta finally catches up with the wind material, the shock interaction between them can further give rise to a detectable radio transient emission on a timescale of tens of days. Discovery of and observations to such dust-affected optical transients and shock-driven radio transients can help to explore the nature of super-Chandrasekhar merger remnants and as well as the density and type ratios of double white dwarf systems, which is beneficial in assessing their gravitational wave contributions.
\end{abstract}

\section{Introduction}
It has been long expected to detect gravitational waves (GWs) in various frequency ranges, in particular, after the first detection of high-frequency GWs in 2015. Close binary white dwarfs (WDs) could contribute a few thousand resolved sources of GWs at frequencies of $\sim 0.01-0.1$ Hz \citep{Nelemans1999, Loren-Aguilar2005} for future GW observatories such as the Laser Interferometer Space Antenna \citep[LISA;][]{Amaro-Seoane2017} and TianQin \citep{Luo2016}. These binaries can also dominate the GW background in the frequency range of $\sim0.01-30$ mHz \citep{Ruiter2010, Yu2010}, which leads to a severe foreground noise for the future GW observations and thus hinders the GW searching for the stellar-mass objects inspiraling into supermassive black holes. In order to precisely assess the GW contributions of binary WDs, it is necessary to constrain the density and type ratios of these systems using electromagnetic observations to them, in particular, to their mergers and merger remnants. Undoubtedly, transient emissions associated with double WD mergers are also one of the most important targets of current time-domain observations.

The outcome of double WD mergers is actually strongly dependent
on the masses of individual WDs and their mass ratio \citep{Iben1984, Iben1996, Webbink1984, Saio1985, Bergeron1991, Nomoto1991, Segretain1997, Han1998, Saio2000}.
Generally, at the initial stage, the less massive WD can be
tidally disrupted by the more massive WD, which remains relatively
undisturbed. For a super-Chandraskehar merger of interest, it could lead to detonation to trigger a type Ia supernova (SN Ia), but likely only if their total mass is higher than $\sim2M_{\odot}$ \citep{Dan2014, Sato2015}. In most cases, the overwhelming majority of material would be bound and only a small mass of $\sim(10^{-3}-10^{-2})M_{\odot}$ can be ejected tidally away from the system. Electromagnetic radiation can, in principle, arise from this tidal outflow due to its cooling and its interaction with the interstellar medium, which are, however, very weak\footnote{\cite{Rueda2018} argued that the thermal emission from the tidal outflow could be detectable due to energy injection from the spin-down of the remnant and fallback accretion. However, due to the existence of the large envelope of the remnant, these energy supplies could be not as efficient as argued.} because of the small ejecta mass and the low-density environment \citep{Moriya2016}. The central remnant of the merger can consist of a core that is contributed by the primary WD, a thermally supported envelope, and a rotationally supported thick accretion disk \citep{Benz1990, Loren-Aguilar2009, Raskin2012, Zhu2013, Dan2014, Becerra2018}. The disk could be long-lived \citep{Yoon2007} but, more probably, will finally blend into the thermal envelope considering that the transport of angular momentum via magnetic stresses can be much faster than the cooling of the remnant \citep{Schwab2012, Shen2012}.

\begin{figure*}
  \centering
  \includegraphics[width=0.75\textwidth]{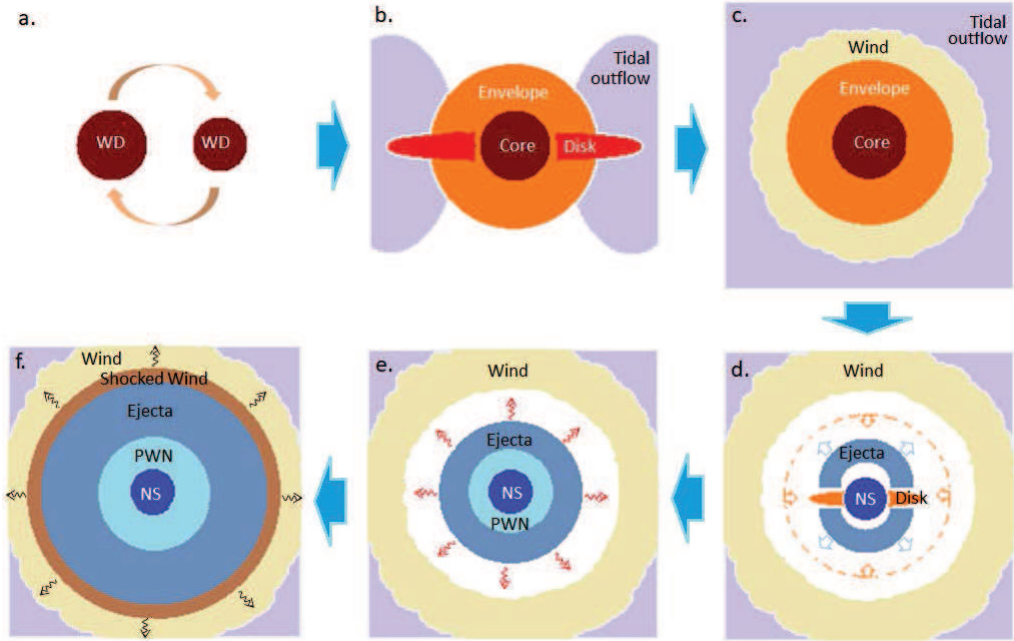}
  \caption{Different evolution stages of a post-merger system including (a) two inspiraling WDs, (b) the initial structure of the merger remnant, (c) the cooling and mass loss of the merger remnant, (d) the collapse to an NS and mass ejection, (e) thermal emission from the ejecta, and (f) nonthermal emission from shocked wind material.}\label{illustration}
\end{figure*}

The long-term evolution of a WD merger remnant, on a thermal timescale of $\sim 10$ kyr, resembles a star evolving from the asymptotic giant branch (AGB) to the planetary nebula stage \citep{Schwab2016}. The
remnant can initially radiate at the Eddington luminosity for a solar
mass object ($\sim10^{38}\rm erg~s^{-1}$), with an effective temperature of $\sim4000-5000$K corresponding to a radius of $\sim 300R_{\odot}$. During this cool and luminous giant phase, the remnant is considered to be able to generate copious amounts of dust and experience significant mass loss \citep{Schwab2016}. This dusty wind can make the remnant obscured and the radiation pressure on the dust can effectively accelerate the wind. Several thousand years later, the mass loss ceases and the wind shell detaches from the remnant surface and evolves into a planetary nebula with a total mass of $\sim 0.1M_{\odot}$. At the same time, the remnant starts to shrink and increase its surface temperature to $\sim 10^5$ K, making it bright in ultraviolet (UV) or soft X-rays. There are three possible fates of the evolution. (i) Most probably, the remnant can finally collapse to form a neutron star (NS), if a carbon flame is ignited off-center and reaches the stellar center that leads to a quiescent conversion of the
remnant to an oxygen¨Cneon composition, to a silicon composition, and to an iron core \citep{Miyaji1980, Nomoto1985, Saio1985, Schwab2015, Wu2018}. (ii) The evolution of the remnant ends at a massive oxygen¨Cneon or silicon WD if its mass finally becomes less than the Chandraskhar critical mass due to a serious mass loss. (iii) A hybrid SN Ia can be triggered, if the carbon flame cannot reach the center \citep{Iben1984, Webbink1984, Wang2018}, which is however unlikely \citep{Lecoanet2016}.

The aftermath of a remnant collapse is the focus of this Letter. The pre- and post-collapse evolution stages are illustrated in Figure \ref{illustration}. After core bounce of the collapse, a neutrino-driven explosion can occur, which is, however, weak\citep{Woosley1992, Dessart2006}. On the contrary, as revealed by \cite{Metzger2009}, a more powerful outflow of a mass of a few times $\sim 0.01M_{\odot}$ can be blown away with a velocity of $\sim0.1c$ from a rotationally supported accretion disk when the disk becomes radiatively inefficient. Here, most of the ejected material can be synthesized into $^{56}$Ni, with respect to a possible electron fraction of $Y_{\rm e}\gtrsim0.5$ due to neutrino irradiation from the proto-NS. Inside such a remnant-collapse-driven (RCD) ejecta, the newborn NS, which probably rotates very quickly, can drive a powerful pulsar wind and then blow a wind bubble. In the next section, we calculate the thermal emission from an RCD ejecta, particularly, by taking into account the influence of a dusty wind shell at large radii. In section 3, we investigate the nonthermal shock emission arising from the interaction of the ejecta with the wind shell. A summary and discussions are given in Section 4.

\section{Thermal optical transient}
Thermal transient emission from an RCD ejecta has been previously studied by \cite{Metzger2009} and \cite{Darbha2010}, where the emission is considered to be powered by the radioactive decay of $^{56}$Ni. Afterwards, \cite{Yu2015} suggested that the spin-down of a newborn NS could enhance the ejecta emission to be as luminous as several times $10^{43}\rm erg~s^{-1}$. With their model, some newly discovered, rapidly evolving, and luminous optical transients can be accounted for as well. Following \cite{Yu2015}, this section is devoted describing the ejecta thermal emission by combining the radioactive and spin-down powers and, in particular, by taking the influence of a surrounding dusty wind into account. The dust effect could make the RCD ejecta emission different from the analogous phenomena arising from mergers of double NSs \citep{Yu2013, Yu2015}.

It is simply considered that both the giant and planetary nebula phases of a merger remnant last for several thousand years, i.e., $t_{\rm w}\sim t_{\rm n}\sim 5$ kyr. Then, the inner radius of the wind shell can be estimated to $r_{\rm w}=v_{\rm w}t_{\rm n}\sim 0.05 \left({v_{\rm w}/ 10{\rm km~s^{-1}}}\right)\left({t_{\rm n}/ 5\rm kyr}\right) {\rm pc}$,
where $v_{\rm w}$ is the speed of wind. In principle, the wind dust grains can be somewhat destroyed by UV emission from the remnant at its final stage before collapse. Following \cite{Waxman2000}, remarkable sublimation of dust grains (e.g., graphite) can take place only if the grain temperature, $T_{\rm g}$ is not too much lower than a critical temperature of $T_{\rm c}\approx 1600{\rm K}[1+0.02\ln (t_{\rm uv,kyr}/a_{-5})]^{-1}$, where $t_{\rm uv,kyr}=t_{\rm uv}/1\rm kyr$ is the UV emission duration and $a_{-5}=a/10^{-5}$ cm is grain size. However, for an injected UV luminosity of $\sim (10^{37}-10^{38})\rm erg~s^{-1}$, the grains at radius $r_{\rm w}$ can only be heated to a temperature of $T_{\rm g}\sim 50$ K. Then, the grain survival time can be estimated to $t_{\rm surv}\sim10^{3}a_{-5}\exp(E_{\rm b}/k_{\rm B}T_{\rm g})\exp(-B/k_{\rm B}T_{\rm c})$ yr, which approaches infinity for $T_{\rm g}\ll T_{\rm c}$, where $k_{\rm B}$ is the Boltzmann constant and $E_{\rm b}\approx 7$ eV is the chemical binding energy per atom. Therefore, dust destruction by the UV emission from the pre-collapse remnant can be neglected safely.

Our calculation of the thermal emission of an RCD ejecta is carried out by using equations presented in Yu et al. (2015) that originated from \cite{Arnett1982} and \cite{Kasen2010}. Specifically, the
bolometric luminosity can be approximately derived from
\begin{equation}
L_{\rm bol}=\left\{
\begin{array}{ll}
{E_{\rm int}c/ (\tau_{\rm opt} R)},&~{\rm for}~\tau_{\rm opt}\geq1,\\
{E_{\rm int}c/  R},&~{\rm for}~\tau_{\rm opt}<1,
\end{array}\right.
\end{equation}
where $E_{\rm int}$ is the internal energy of the ejecta that is considered to be radiation-dominated, $\tau_{\rm opt}=3\kappa_{\rm opt} M_{\rm ej}/4\pi R^2$
is the optical depth for optical emission, $\kappa_{\rm opt}$ is the corresponding opacity, and $M_{\rm ej}$ and $R$ are the mass and outmost radius of the ejecta, respectively. The evolution of the internal energy is
determined by the energy conservation law as
\begin{eqnarray}
\frac{ d E_{\rm int}}{d t} =  L_{\rm h} - L_{\rm bol}- P \frac{ d
V}{d t},\label{Eint1}
\end{eqnarray}
where $L_{\rm h}$ is the heating power that is contributed by both radioactivity and NS spin-down, the work $PdV=4\pi R^2Pv_{\rm ej}dt$
represents the adiabatic cooling, and $P$ and $v_{\rm ej}$ are the pressure and velocity of the ejecta, respectively.

\begin{figure}
\begin{center}
\includegraphics[scale=0.3]{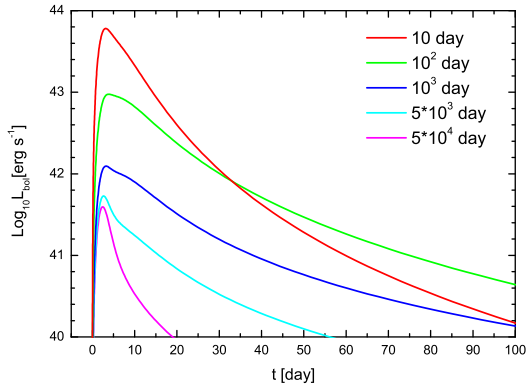}
\caption{Bolometric light curves of ejecta thermal emission for different spin-down timescales as labeled, where the influence of a dusty wind shell is not taken into account.
}\label{Lbol}
\end{center}
\end{figure}

The heating rates due to radioactivity and NS spin-down are given by
\begin{eqnarray}
L_{\rm ra}=f_{\gamma}\left[L_{\rm Ni}
\exp\left(-{t\over t_{\rm Ni}}\right)+L_{\rm Co}
\exp\left(-{t\over t_{\rm Co}}\right)\right],
\label{Lra}
\end{eqnarray}
and
\begin{eqnarray}
L_{\rm sd}=f_{\rm X}{E_{\rm rot}\over t_{\rm sd}}\left(1+{t\over t_{\rm sd}}\right)^{-2},
\label{Lsd}
\end{eqnarray}
respectively, where $L_{\rm Ni}=8.0\times
10^{43}({M_{\rm Ni}/ M_{\odot}})~ \rm erg\
s^{-1}$ and $L_{\rm Co}=1.4\times
10^{43}({M_{\rm Ni}/ M_{\odot}})~ \rm erg\
s^{-1}$ with $M_{\rm Ni}$ being the initial mass of $^{56}$Ni, $t_{\rm Ni}=8.8$ days and $t_{\rm Co}=111.3$ days are the mean lifetimes of the nuclei, $E_{\rm rot}=10^{51}(P_{\rm i}/5{\rm ms})^{-2} \rm erg$ is the rotational energy of the NS, $t_{\rm sd}=1500(B_{\rm p}/10^{13}{\rm G})^{-2}(P_{\rm i}/5\rm ms)^2$ day is the spin-down timescale, and $B_{\rm p}$ and $P_{\rm i}$ are the dipolar magnetic field strength and the initial spin period, respectively. Here the NS spin-down is considered to be dominated by magnetic dipole radiation. While the gamma-rays from
radioactive decay of $^{56}$Ni and $^{56}$Co are around 2 MeV, the energy released from an NS is considered to be primarily in the form of soft to hard X-rays\footnote{As an ultra-relativistic pulsar wind initially collides with an RCD ejecta, a forward shock wave can form and propagate into the ejecta. Nevertheless, due to the high optical depth of the ejecta, nonthermal emission could not be produced by this shock, unless the NS spin-down is significantly delayed and the ejecta has moved to very large radii \citep{Piro2013}. Instead, thermal emission could appear at a very early time ($\sim 10^3$ s) when the shock breaks out from the ejecta \citep{Li2016}.} \citep{Yu2007, Metzger2014}, which are generated by electrons and positrons in a pulsar wind nebula (PWN). Therefore, the coefficients
$f_{\gamma}\equiv1-\exp(-\tau_{\gamma})$ in Equation (\ref{Lra}) and $f_{\rm X}\equiv \xi[1-\exp(-\tau_{\rm X})]$ in Equation (\ref{Lsd}) represent the  respective thermalization efficiency of the injected energy, where $\tau_{\gamma}$ and $\tau_{\rm X}$ are the optical depths for gamma-rays and X-rays, respectively. Finally, the extra factor $\xi\sim0.1$ is introduced in $f_{\rm X}$ since a remarkable fraction of X-rays emitted from the PWN can be reflected into the PWN by the ejecta \citep{Metzger2014}. The opacity of an RCD ejecta must be highly
energy dependent. As shown in Figure 8 of Kotera et al. (2013), for a normal SN ejecta, the opacity can vary from $\sim10^6\rm cm^2g^{-1}$
for $\sim 100$ eV to $\sim0.01\rm cm^2g^{-1}$ above $\sim 10$ MeV. To be specific, the opacity in the energy range of $\sim0.1-100$
keV could be dominated by photoelectric absorption, while the Compton
scattering and pair production processes determine the opacity above $\sim100$ keV and above $\sim10$
MeV, respectively. By according to this result, we analogically take $\kappa_{\rm X}=1.0\rm~ cm^2~g^{-1}$ and $\kappa_{\gamma}\sim 0.03\rm~ cm^2~g^{-1}$ in our calculations. Then, from equations $\tau_{\rm X}=1$ and $\tau_\gamma=1$, the leakage times of X-rays and gamma-rays can be derived to $t_{\rm lk,X}=12(\kappa_{\rm X}/1.0{\rm cm^2g^{-1}})^{1/2}(M_{\rm ej}/0.02M_{\odot})^{1/2}(v_{\rm ej}/0.1c)^{-1}$ days and $t_{\rm lk,\gamma}=2(\kappa_{\gamma}/0.03{\rm cm^2g^{-1}})^{1/2}(M_{\rm ej}/0.02M_{\odot})^{1/2}(v_{\rm ej}/0.1c)^{-1}$ days.
\begin{table*}
\centering
\renewcommand{\arraystretch}{2.0}\caption{Fiducial model parameters}
\begin{tabular}{ccc|cccccc|ccccc}
 \hline
 \hline
$E_{\rm rot}$&$t_{\rm sd}$&$\xi$&$M_{\rm ej}$& $M_{\rm Ni}$&$v_{\rm ej,i}$& $\kappa_{\rm opt}$& $\kappa_{\rm X}$& $\kappa_{\gamma}$&$\kappa_{\rm dust}$&$r_{\rm w}$
\\\hline
$10^{51}$ erg &1000 days&0.1&$0.02M_{\odot}$& $0.01M_{\odot}$&$0.1c$& $0.2\rm~ cm^2~g^{-1}$& $1.0\rm~ cm^2~g^{-1}$& $0.03\rm~ cm^2~g^{-1}$&$300\rm~ cm^2~g^{-1}$&0.05 pc

\\
\hline \hline
\end{tabular}
\end{table*}

The bolometric light curves of the ejecta thermal emission are presented in Figure \ref{Lbol} for different spin-down timescales of the NS. Throughout this Letter, the model parameters are taken as their fiducial values listed in Table 1 unless otherwise specified. As is well known, the peak time of the bolometric light curves is basically determined by photon diffusion on a timescale of $t_{\rm d}=(3\kappa_{\rm opt} M_{\rm ej}/4\pi v_{\rm ej}c)^{1/2}=1.7(\kappa_{\rm opt}/0.2{\rm cm^2g^{-1}})^{1/2}(M_{\rm ej}/0.02M_{\odot})^{1/2}(v_{\rm ej}/0.1c)^{-1/2}$ days. Therefore, according to the Arnett law \citep{Arnett1982} and comparing the values of $L_{\rm ra}(t_{\rm d})$ and $L_{\rm sd}(t_{\rm d})$, it can be found that the spin-down power can dominate the thermal emission as long as $t_{\rm sd}\lesssim 1500(\xi/0.1)(P_{\rm i}/5\rm ms)^{-2}(M_{\rm Ni}/0.01M_{\odot})^{-1}$ days, which corresponds to $B_{\rm p}\gtrsim10^{13}$ G. In particular, if the NS can become a magnetar of a field much higher than $\sim4\times10^{13}$ G, then the resultant emission can arrive at a peak luminosity much higher than $\sim10^{43}\rm erg~s^{-1}$, which can be employed to explain some unusual luminous and rapidly evolving optical transient events \citep{Drout2014, Vinko2015, Yu2015, Pursiainen2018, Rest2018}.

\begin{figure}
\begin{center}
\includegraphics[scale=0.3]{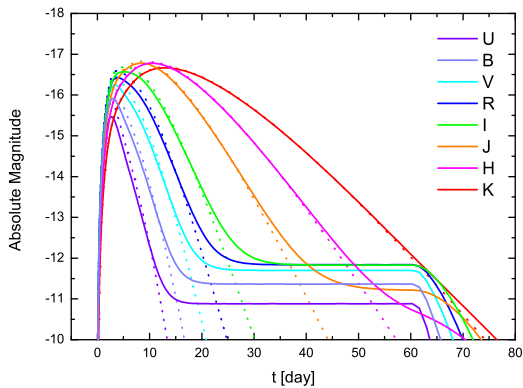}
\caption{Multiband light curves affected (solid lines) and unaffected (dotted lines) by a dusty wind shell.
}\label{Lcol1}
\end{center}
\end{figure}

\begin{figure}
\begin{center}
\includegraphics[scale=0.3]{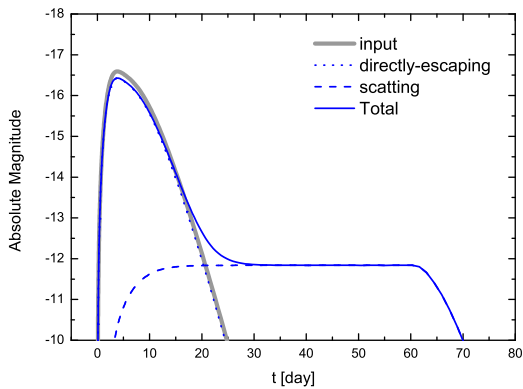}
\caption{$R$-band light curve consisting of a directly escaping emission component and a scattering component. The input emission is presented by the gray solid line.
}\label{dusteffect}
\end{center}
\end{figure}

\begin{figure}
\begin{center}
\includegraphics[scale=0.3]{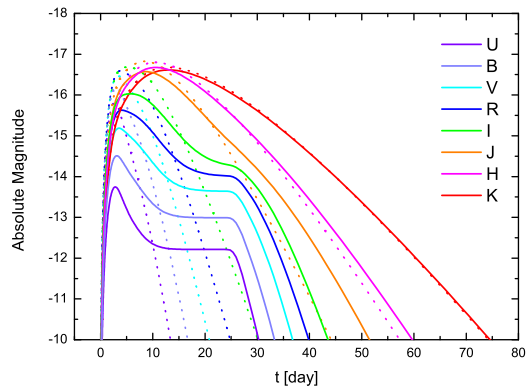}
\caption{Same as that in Figure 3, but for $r_{\rm w}=0.02$ pc.}\label{Lcol2}
\end{center}
\end{figure}

Due to the existence of the dusty wind shell at large radii, the ejecta thermal emission can be somewhat absorbed and scattered. For an input luminosity, $L_{\lambda}(t)$, described by a blackbody spectrum with an effective temperature of $T_{\rm eff}(t)=[L_{\rm bol}(t)/4\pi R(t)^2\sigma]^{1/4}$, where $\sigma$ is the Stephan¨CBoltzmann constant, the luminosities of the directly escaping emission component and the scattering component can be calculated by
\begin{eqnarray}
L_{\lambda}^{\rm dir}(t)=L_{\lambda}(t)10^{-0.4(A_{\lambda}+G_{\lambda})}
\end{eqnarray}
and
\begin{eqnarray}
L_{\lambda}^{\rm scat}(t)&=&\int_0^{\pi/2}L_{\lambda}[t-\delta t(\theta)]\sin \theta d\theta\nonumber\\
&&\times10^{-0.4A_{\lambda}}(1-10^{-0.4G_{\lambda}}),\label{scatt}
\end{eqnarray}
respectively. Here, for the scattering emission, the different time delays for different scattering angles are determined by $\delta t(\theta)=r_{\rm w}(1-\cos\theta)/c$. For the extinction $A_{\lambda}$ and as well as $G_{\lambda}$, empirical numerical descriptions can be obtained from \cite{Kochanek2012}, as tabulated in Tables 3 and 4 therein. Both values of $A_{\lambda}$ and $G_{\lambda}$ depend on the visual optical depth of the dust shell, which can be estimated by
\begin{eqnarray}
\tau_{\rm dust}&=&{\kappa_{\rm dust}M_{\rm w}\over 4\pi r_{\rm w}^2}\nonumber\\
&=&0.2\left({\kappa_{\rm dust}\over 300 \rm cm^2g^{-1}}\right)\left({M_{\rm w}\over 0.1M_{\odot}}\right)\left({r_{\rm w}\over 0.05\rm pc}\right)^{-2},
\end{eqnarray}
where $\kappa_{\rm dust}$ is the dust opacity for optical emission and $M_{\rm w}$ is the total mass of the dusty wind.

In Figure \ref{Lcol1}, we plot the dust-affected multiband light curves of the ejecta emission. The constitute of the directly escaping and scattering emission components in a light curve is shown in Figure \ref{dusteffect}. Due to the dust scattering, an obvious plateau appears in the light curves. The duration of the plateau emission is determined by $r_{\rm w}/c$. A relatively small $r_{\rm w}$ can shorten the scattering delay time and increase the dust optical depth of the wind shell. By arbitrarily taking $r_{\rm w}=0.02$ pc, we re-plot the multiband light curves in Figure \ref{Lcol2}. In this case, the peak emission is obviously suppressed by the increased optical depth, whereas the scattering plateau emission is significantly enhanced, particularly in UVBRI bands. Finally, by considering of the emission absorption, it is expected that the heated wind material could, in principle, contribute a detectable infrared emission if the $\tau_{\rm dust}$ is high enough.

\section{Nonthermal shock emission}
After the thermal transient emission, the RCD ejecta can finally catch up with the preceding wind shell and drive a shock propagating into the wind material. As a result, nonthermal shock emission can be generated. In this section, the time, $t$, is redefined by setting the time to zero at the collision between the ejecta and wind, which is delayed by several years (i.e., $r_{\rm w}/v_{\rm ej}$) from the thermal optical transient.

The dynamical evolution of the shock can be determined by the energy conservation law as $E_{\rm ej}=(1/2)M_{\rm ej}v_{\rm s}^2+M_{\rm sw}v_{\rm s}^2$, where $E_{\rm ej}$ is the final kinetic energy of the RCD ejecta and the term $M_{\rm sw}v_{\rm s}^2$ includes both the bulk kinetic and internal energy of the wind material of a mass of $M_{\rm sw}$ that is swept up by the shock. The density profile of the wind material can be written as $\rho(r)=({M}_{\rm w}/ 4\pi v_{\rm w}t_{\rm w })r^{-2}\equiv Ar^{-2}$,
where $t_{\rm w}$ is the duration of the mass-losing AGB stage of the merger remnant. Then, the mass of shocked material can be obtained by
\begin{eqnarray}
M_{\rm sw}=4\pi\int_{r_{\rm w}}^{r_{\rm sf}}\rho(r)r^2dr=4\pi A (r_{\rm sf}-r_{\rm w}),
\end{eqnarray}
where $r_{\rm sf}$ is the radius of shock front. By substituting the above expression into $v_{\rm s}=[{2E_{\rm ej}/( M_{\rm ej}+2M_{\rm sw}})]^{1/2}$, we get
\begin{eqnarray}
v_{\rm s}=v_{\rm ej}\left[1+{(r_{\rm sf}-r_{\rm w})\over r_{*}}\right]^{-1/2},\label{vsr}
\end{eqnarray}
where $r_{*}\equiv M_{\rm ej}/(8\pi A)$.
The combination of the above equation with ${dr_{\rm sf}/ dt}=v_{\rm sf}$ and $v_{\rm sf}=(4/3)v_{\rm s}$ (Lang 1980) yields
\begin{eqnarray}
r_{\rm sf}(t)=r_{\rm w}+r_{*}\left[\left(1+{t\over t_{\rm dec}}\right)^{2/3}-1\right]\label{rsf}
\end{eqnarray}
and
\begin{eqnarray}
v_{\rm s}(t)=v_{\rm ej}\left(1+{t\over t_{\rm dec}}\right)^{-1/3},\label{vst}
\end{eqnarray}
where the deceleration timescale reads $t_{\rm dec}\equiv r_{*}/(2v_{\rm ej})$. By assuming the outmost radius of the wind shell to be about $2r_{\rm w}$, the timescale on which the shock crosses the shell can be derived from Equation (\ref{rsf}) to
\begin{eqnarray}
t_{\rm cr}&=&{r_{\rm w}^{3/2}\over2r_{*}^{1/2}v_{\rm ej}}\nonumber\\
&=&2.6\left({r_{\rm w}\over 0.05\rm pc}\right)\left({M_{\rm ej}\over 0.02 M_{\odot}}\right)^{-1/2}\left({v_{\rm ej}\over 0.1c}\right)^{-1}\rm yr.
\end{eqnarray}
After this crossing time, the shock would rush into the material that is tidally ejected a few ten thousand years ago. However, the shock emission at that time could become very weak.

Following \cite{Huang2003}, the shock-accelerated
electrons are considered to initially distribute as a power law with their kinetic
energy as ${dN_{\rm e}/ d{\gamma}} \propto ({\gamma}-1)^{-p}$ for ${\gamma}\geq{\gamma}_m$.
The minimum Lorentz factor of the electrons reads $\gamma_{\rm m}=\epsilon_{\rm e}g_p{(m_{\rm p}/
2m_{\rm e})}(v_{\rm s}/ c)^2+1$, where $\epsilon_{\rm e}$ is the energy equipartition factor of the electrons, $g_p=(p-2)/(p-1)$, and $m_{\rm p}$ and $m_{\rm e}$ are the proton and electron masses, respectively. According to this distribution, the number of relativistic electrons of Lorentz factors higher than $\sim2$ can be given by $N_{\rm e,rel}\thickapprox (\gamma_{\rm m}-1)^{(p-1)}M_{\rm sw}/m_{\rm p}$. We calculate the synchrotron emission of these relativistic electrons using an analytical method widely used in gamma-ray burst researches \citep{Sari1998, Panaitescu2000}. First, the strength of the magnetic fields in the shocked region can be expressed by $B=(4\pi\epsilon_{\rm B}\rho_{\rm s} v_{\rm s}^2)^{1/2}$ by introducing an equipartition factor, $\epsilon_{\rm B}$, where $\rho_{\rm s}=4\rho(r_{\rm sf})$. Then, a cooling Lorentz factor of the electrons is given by $\gamma_{\rm c}={6\pi m_{\rm e} c/(\sigma_{\rm T}B^2t)}$, where $\sigma_{\rm T}$ is the Thomson cross section. Furthermore, three characteristic frequencies are defined as follows: peak frequency, $\nu_{\rm p}\sim{2eB/(\pi m_{\rm e}c)}$, that corresponds to $\gamma\sim2$, 
cooling frequency, $\nu_{\rm c}={eB\gamma_{\rm c}^2/(2\pi m_{\rm e}c)}$, and self-absorption frequency $\nu_{\rm a}=[5eN_{\rm e,rel}/(128\pi r_{\rm sf}^2B)]^{2/(p+4)}\nu_{p}$, where $e$ is the electron charge.
For the available ranges of the model parameter values, we always have $\nu_{\rm p}<\nu_{\rm a}<\nu_{\rm c}$. Therefore, the following spectrum is employed to calculate the shock emission for a frequency $\nu$:
\begin{equation}
L_{\nu}^{\rm syn}=L_{\nu,\max}^{\rm syn}\times\left\{
\begin{array}{ll}
\left({\nu_{\rm a}\over\nu_{\rm p}}\right)^{-{(p-1)\over2}}\left({\nu_{\rm }\over\nu_{\rm a}}\right)^{{5\over2}},&\nu_{\rm p}<\nu<\nu_{\rm a};\\
\left({\nu\over\nu_{\rm p}}\right)^{-{(p-1)\over2}},&\nu_{\rm a}<\nu<\nu_{\rm c};\\
\left({\nu_c\over\nu_{\rm p}}\right)^{-{(p-1)\over2}}\left({\nu\over\nu_{\rm c}}\right)^{-{p\over2}},&\nu>\nu_{\rm c},\\
\end{array}\right.
\end{equation}
where the peak spectral power is given by $L_{\nu,
\max}^{\rm syn}= N_{\rm e,rel}{m_{\rm e}c^2\sigma_{\rm T}B/ (3e)}$.

\begin{figure}
\begin{center}
\includegraphics[scale=0.3]{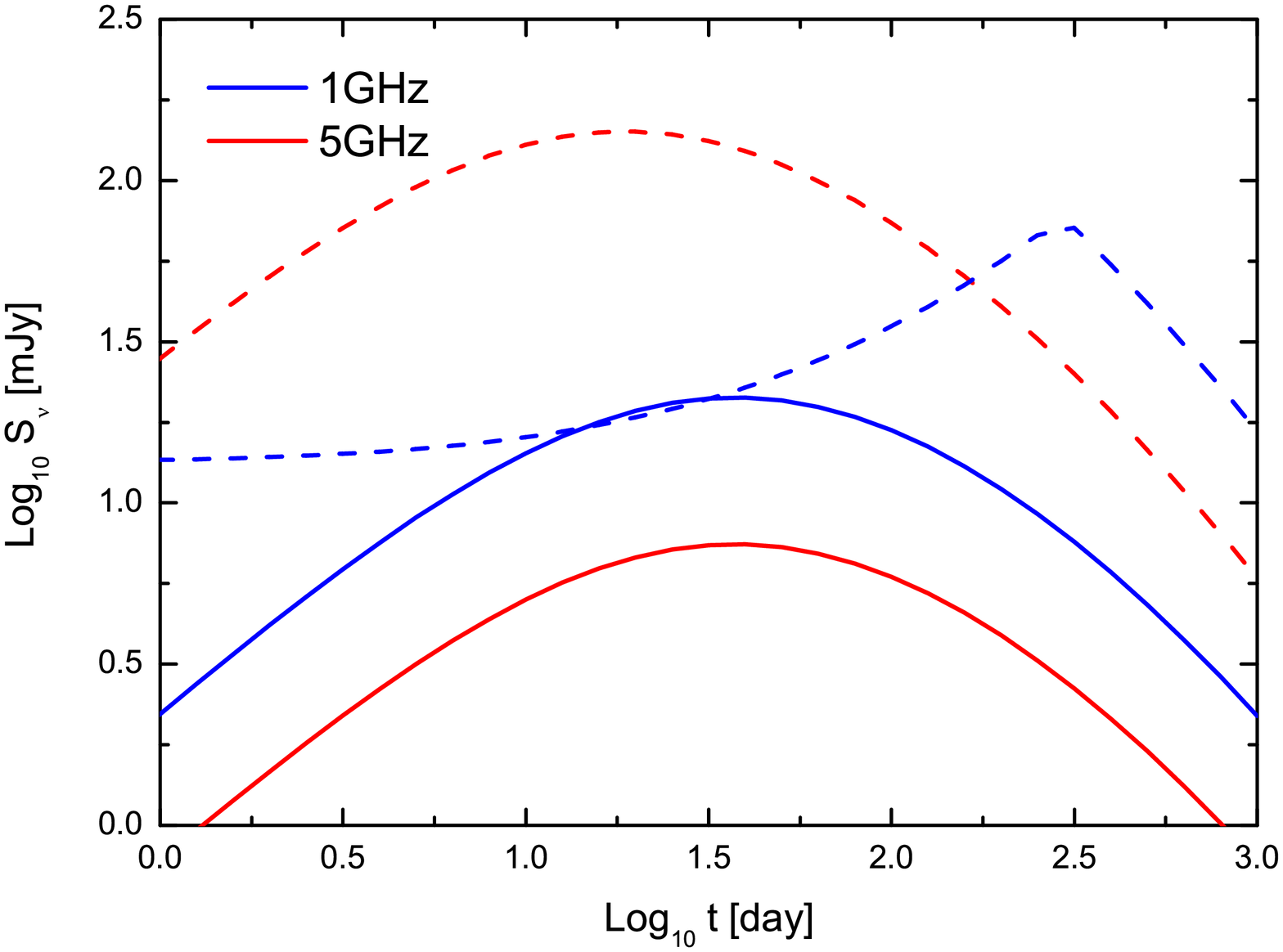}
\caption{Radio emission arising from the shock interaction between an RCD ejecta and preceding wind material, where a distance of 100 Mpc is taken. The solid and dashed lines are given for  $v_{\rm ej}=0.1c$ and $0.2c$, respectively. The shock parameters are taken as $p=2.3$, $\epsilon_{\rm e}=0.1$, and $\epsilon_{\rm B}=0.01$. The zero point of time $t$ is set at the moment when the RCD ejecta reaches the radius $r_{\rm w}$, which is delayed by several years from the NS formation. }\label{LCnth}
\end{center}
\end{figure}

Since the shock emission in the optical and X-ray bands is too weak to be detected, in Figure \ref{LCnth} we only present the radio light curves in the 1 and 5 GHz bands. For an example distance of $100$ Mpc, the radio flux $S_{\nu}$ can range from several mJy to a few ten mJy with a peak value of
\begin{eqnarray}
S_{\nu,\rm p}^{\rm}&\approx&23{\rm mJy}\left({\nu\over 1{\rm GHz}}\right)^{1-p\over2}\left({d\over 100{\rm Mpc}}\right)^{-2}\left({M_{\rm w}\over 0.1M_{\odot}}\right)^{p+1\over4}\nonumber\\
&&\times\left({r_{\rm w}\over 0.05{\rm pc}}\right)^{-{(p+1)\over2}}\left({v_{\rm w}\over 10{\rm km~s^{-1}}}\right)^{-{(p+1)\over4}}\left({t_{\rm w}\over 5\rm kyr}\right)^{-{(p+1)\over4}}\nonumber\\
&&\times\left({M_{\rm ej}\over 0.02M_{\odot}}\right)\left({v_{\rm ej}\over 0.1c}\right)^{{5p-3\over2}}\left({\epsilon_{\rm e}\over 0.1}\right)^{p-1}\left({\epsilon_{\rm B}\over 0.01}\right)^{p+1\over4},
\end{eqnarray}
if the observational frequency, $\nu$, is higher than the self-absorption frequency as
\begin{eqnarray}
\nu_{\rm a}(t_{\rm p})&=&0.4{\rm GHz}\left({M_{\rm w}\over 0.1M_{\odot}}\right)^{p+2\over2(p+4)}\left({r_{\rm w}\over 0.05{\rm pc}}\right)^{-{(p+6)\over(p+4)}}\nonumber\\
&&\times\left({t_{\rm w}\over 5\rm kyr}\right)^{-{(p+2)\over2(p+4)}}\left({M_{\rm ej}\over 0.02M_{\odot}}\right)^{2\over p+4}\left({v_{\rm ej}\over 0.1c}\right)^{{5p-2\over p+4}}\nonumber\\
&&\times\left({\epsilon_{\rm e}\over 0.1}\right)^{2(p-1)\over p+4}\left({\epsilon_{\rm B}\over 0.01}\right)^{p+2\over2(p+4)},
\end{eqnarray}
where the coefficient is obtained for $p=2.3$. The peak time $t_{\rm p}$ of this radio transient is determined by the shock deceleration timescale as
\begin{eqnarray}
t_{\rm dec}&=&30{\rm day}\left({M_{\rm ej}\over 0.02M_{\odot}}\right)\left({v_{\rm ej}\over 0.1c}\right)^{-1}\nonumber\\
&&\times\left({M_{\rm w}\over 0.1M_{\odot}}\right)^{-1}\left({v_{\rm w}\over 10{\rm km~s^{-1}}}\right)\left({t_{\rm w}\over 5\rm kyr}\right).
\end{eqnarray}

Finally, we would like to point out that the majority of the spin-down energy of an NS could be trapped within a PWN because of the X-ray reflection by the ejecta base. Therefore, as long as the reflection is efficient, the high pressure of the PWN can effectively compress and accelerate the RCD ejecta. Then, the final kinetic energy of the ejecta would be close to $(1-\xi)E_{\rm rot}$, i.e., $v_{\rm ej}\rightarrow 0.2c$, when it collides with the wind shell. Nevertheless, this kinetic energy is primarily injected during the catching-up process, since the catching-up time is much longer than the dynamical time of the shock. So, the time dependence of $E_{\rm ej}$ can still be ignored in dynamical equations (\ref{vsr}) and (\ref{vst}). By tentatively taking $v_{\rm ej}=0.2c$, we show in Figure \ref{LCnth} that the higher frequency (5 GHz) emission can be increased significantly to $\sim 100$ mJy. However, the lower frequency (1 GHz) emission would be seriously suppressed by the enhanced synchrotron self-absorption and, correspondingly, its emission peak can be delayed much from the deceleration time. Furthermore, if the NS actually can rotate more quickly as considered, then the ejecta can even be accelerated to a mild relativistic velocity and the shock emission is shifted to optical or X-ray bands (see \cite{Gao2013} for an analogical consideration for the double NS merger case).

\section{Summary and discussions}
Mergers of double WDs are very competitive origins of SNe Ia, which is strongly supported by the observational distribution of the delayed times of these SNe \citep[e.g.][]{Wang2012,Maoz2014}. Nevertheless, on an alternative channel, these mergers could probably lead to the formation of a massive WD. If this post-merger WD is super-Chandrasekhar, it can finally, after a long-term cooling, collapse into an NS accompanying with an ejection of a $^{56}$Ni-rich outflow. On the one hand, a day-scale thermal transient can be produced by the ejecta due to heating by radioactivity and NS spin-down. The peak luminosity of this optical transient can range from $\sim10^{41}\rm erg~s^{-1}$ to several times $\sim10^{43}\rm erg~s^{-1}$. Moreover, this optical emission could be substantially affected by a surrounding dusty wind shell at large radii, which was produced by the merger remnant several thousand years ago before the NS formation. This dust shell can somewhat suppress the peak emission and contribute a scattering emission plateau in light curves, making these transients different from analogous phenomena called kilonovae/mergernovae due to double NS mergers. On the other hand, a month-scale nonthermal radio transient can be produced by the shock interaction between the ejecta and the wind material, which takes place about several years after the thermal optical transient. These optical and radio transients provide an available way to discover and identify the collapse of super-Chandrasekhar remnants, to probe the nature and evolution of these remnants, and to constrain the ratio of these collapse events to SN Ia events.

It is worth mentioning that, although a significant fraction of the envelope of a merger remnant can be lost through stellar wind, the remaining envelope might still keep some matter from falling onto the central proto-NS, if the total mass of the remnant is much higher than the Chandrasekhar mass. In this particular case, after collapse, the RCD ejecta can quickly collide with the remaining envelope. Then, the ejecta kinetic energy can be consumed to power the thermal transient emission additionally, as previously studied by \cite{Metzger2009} and \cite{Brooks2017}.

The local rate of SNe Ia has been measured to be around
$(3.01 \pm 0.062) \times 10^4 \rm Gpc^{-3} yr^{-1}$ \citep{Li2011}. Therefore, it is believed that the total event rate of WD mergers could be somewhat higher than or, at least, not much lower than $\sim10^4 \rm Gpc^{-3} yr^{-1}$, although SNe Ia can also be contributed by a single WD accreting from its companion star. In any case, this order of magnitude of the WD merger rate can be well accounted for by some simulations \citep[e.g.][]{Badenes2012}. On the one hand, the high merger rate makes it possible to discover these predicted optical and radio transients in the future. On the other hand, these future transient observations are eagerly expected to confirm the above rate estimation and to constrain the ratio of different types of WD mergers by combining with observations of SNe Ia.


\acknowledgements
This work is supported by the National Natural Science
Foundation of China (grant No. 11473008, 11822302, 11833003, and 11873085).

\end{document}